\documentclass[twocolumn, linenumber]{aastex63}

\newcommand{\msun}{\ensuremath{{\rm M}_\odot}}
\newcommand{\lowerPISN}{55\msun}
\newcommand{\upperPISN}{120\msun}

\newcommand{\NC}{NC21}

\newcommand{\citeg}[1]{\citep[e.g.,][]{#1}}

\graphicspath{{./}{figures/}}

\usepackage{multirow}
\usepackage{lineno}
\usepackage{amsmath}

\begin{document}

\title{Poking Holes: Looking for Gaps in LIGO/Virgo's Black Hole Population}

\author{Bruce Edelman}
\email{bedelman@uoregon.edu} 
\affiliation{Institute  for  Fundamental  Science, Department of Physics, University of Oregon, Eugene, OR 97403, USA}

\author{Zoheyr Doctor}
%\email{zdoctor@uoregon.edu}
\affiliation{Institute  for  Fundamental  Science, Department of Physics, University of Oregon, Eugene, OR 97403, USA}

\author{Ben Farr}
%\email{bfarr@uoregon.edu}
\affiliation{Institute  for  Fundamental  Science, Department of Physics, University of Oregon, Eugene, OR 97403, USA}

\begin{abstract}
Stellar evolution models predict the existence of a gap in the black hole mass spectrum from $\sim\lowerPISN-\upperPISN$ due to pair-instability supernovae (PISNe). We investigate the possible existence of such an ``upper" mass gap in the second gravitational wave transient catalog (GWTC-2) by hierarchically modeling the astrophysical distribution of black hole masses. We extend the \textsc{Truncated} and \textsc{Powerlaw+Peak} mass distribution families to allow for an explicit gap in the mass distribution, and apply the extended models to GWTC-2. We find that with the \textsc{Truncated} model there is mild evidence favoring an upper mass gap with log Bayes Factor $\ln \mathcal{B} = 2.79$, inferring the lower and upper bounds at $56.12_{-4.38}^{+7.54} \msun$, and $103.74_{-6.32}^{+17.01}\msun$ respectively. When using the \textsc{Powerlaw+Peak} model, we find no preference for the gap. When imposing tighter priors on the gap bounds centered on the expected PISNe gap bounds, the log Bayes factors in favor of a gap mildly increase. These results are however contingent on the parameter inference for the most massive binary, GW190521, for which follow up analyses showed the source may be an intermediate mass ratio merger that has component masses straddling the gap. Using the GW190521 posterior samples from the analysis in \citet{Nitz_2021}, we find an increase in Bayes factors in favor of the gap. However, the overall conclusions are unchanged: There is no preference for a gap when using the \textsc{Powerlaw+Peak} model. This work paves the way for constraining the physics of pair-instability and pulsational pair-instability supernovae and high-mass black hole formation. 

\end{abstract}

\section{Introduction} \label{sec:intro}

With the recent release of its second gravitational wave transient catalog (GWTC-2), the LIGO/Virgo collaboration (LVC) has now detected 50 gravitational wave (GW) events since the start of the advanced detector era, at least 46 of which came from binary black hole (BBH) systems \citep{aLIGO, aVIRGO, GWTC1, gwtc2}. GWTC-2 therefore provides a rich data set to infer properties of the astrophysical population of stellar mass black holes \citep{o1o2_pop, o3a_pop}. A robustly predicted feature that we can look for, specifically in the black hole (BH) mass distribution, is the theorized upper mass gap produced from effects due pair instability supernovae (PISNe) which precludes formation of BHs with masses $\sim\lowerPISN-\upperPISN$ from stellar collapse. 

\begin{figure*} [ht]
    \centering
    \includegraphics[width=0.9\textwidth]{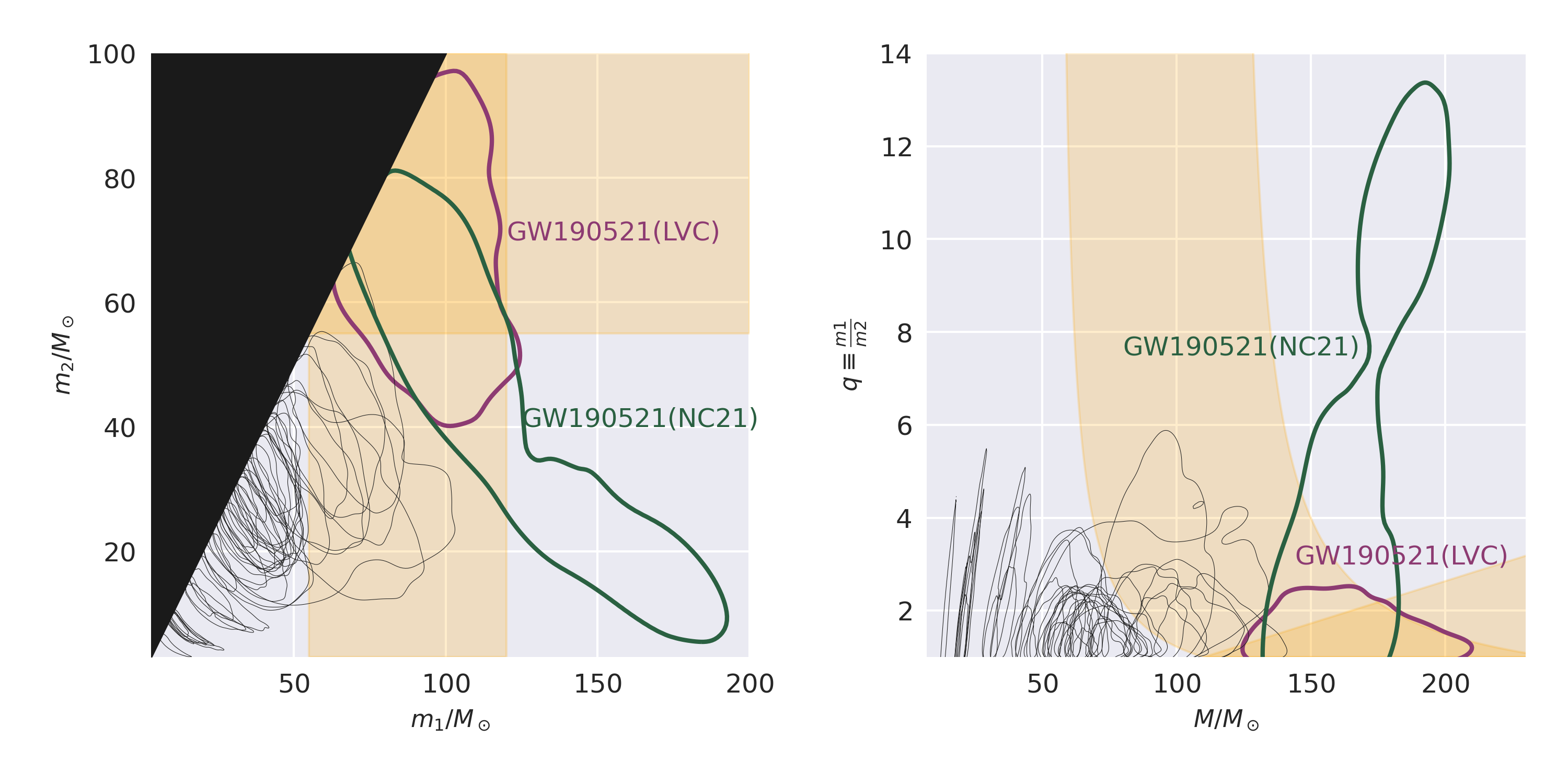}
    \caption{90\% credible level contour of the posterior samples for each of the 46 BBH mergers in GWTC-2. We show both sets of posterior samples for the highest mass event, GW190521, from \citep{Nitz_2021} (green) and from the LVC analysis (purple). Posterior samples from \citet{Nitz_2021} have been re-weighted to the same priors as the LVC analyses. The approximate expected region ($\sim \lowerPISN-\upperPISN$) of the PISNe mass gap is highlighted in orange.}
    \label{fig:GWTC2_contours}
\end{figure*}

Stellar evolution simulations show that stars with core masses from $\sim40-135\msun$ undergo PISNe in which the highly energetic gamma rays produced in the core can collide with atomic nuclei and produce electron-positron pairs \citep{PISN_Woosley}. This production process absorbs energy that was previously counteracting the gravitational pressure causing the core to contract. Heavier core stars in the $\sim65-135\msun$ range ignite oxygen leading to an unstable thermonuclear explosion which leaves behind no compact remnant \citep{Heger_2002, Heger_2003}. Lighter stars with core masses $\sim40-65\msun$ can temporarily stabilize themselves after the ignition and thus go through a series of pulsations (PPISN) that shed large amounts of mass with each pulse \citep{Woosley_2017}. This continues until the mass of the star is too light to pair-produce, leaving the star to undergo a normal core-collapse supernova (CCSN) that leaves behind remnant black holes of masses $\lesssim\lowerPISN$ \citep{Woosley_2019}. Even more massive low metallicity stars can bypass PISN completely and possibly form intermediate-mass black holes (IMBHs) with masses $>\upperPISN$ \citep{Spera_2017}. While simulations consistently predict the existence of this proposed mass gap in the distribution of binary black hole masses, the precise locations of the lower and upper boundaries remain uncertain \citep{Belczynski_2016,Farmer_2019, Stevenson_2019, Farmer_2020, Belczynski_2020, Vink_2021}. Analyses of GW data from the first and second Advanced LIGO-Virgo observing runs put constraints on the lower bound of this mass gap by modeling the black hole primary mass distribution as a powerlaw with a sharp high-mass cutoff, and found support for this lower edge at $\sim45\msun$ \citep{o1o2_pop, Fishbach_2017, Talbot_2018}. While the GWTC-2 catalog is consistent with 97\% of observed BBH primary masses lying below $45\msun$, population analyses using GWTC-2 and parameterized toy models find that there is less support for a sharp cutoff and instead preference for a shallow tail to larger masses \citep{o3a_pop}. This shallow tail extending into the supposedly forbidden range of masses is primarily driven by the few GW detections that have posterior support in the mass gap, most notably GW190521 \citep{190521_disc, 190521_astro_imp}. This leads to the possibility that there could be a separate population of black holes contributing in that mass range, formed through some other mechanism. One possibility is that remnants of previous black hole mergers undergo subsequent ``hierarchical" mergers which, in dynamical environments such as globular clusters or active galactic nuclei, can contribute a significant fraction to the overall rate of mergers \citep{Romero_Shaw_2020, Kimball_genealogy, Doctor_2020, gayathri2020gw190521, Farrell_2020, Secunda_2020, McKernan_2020}. For example, \citet{kimball2020evidence} finds evidence that GWTC-2 includes at least one merger with a second generation component under certain assumptions about the 1st generation black hole mass distribution. 

The hierarchical merger scenario is not the only explanation for high mass events like GW190521, though. Other recent work has looked more closely at the inferred source parameters of GW190521. Using gravitational waveforms with quasi-circular black hole inspirals and a standard ``agnostic" prior, the LVC found the source of GW190521 to have component source-frame masses within the theorized bounds of the mass gap with $m_1^{\mathrm{src}} = 85_{-14}^{+21}\msun$ and $m_2^{\mathrm{src}} = 66_{-18}^{+17}\msun$ \citep{190521_disc, 190521_astro_imp}. However, other waveform models and priors lead to other interpretations of this event. For example, the source of GW190521 could have had a highly eccentric orbit, been a head-on merger, or been subject to new physics allowing formation within the PISNe mass gap \citep{Romero_Shaw_2020, gayathri2020gw190521, bustillo2020confusing, Sakstein_2020, cruzosorio2021gw190521}. Alternatively, assuming the secondary component of GW190521 comes from the same prior distribution of secondaries as other events, there is more support for the components of GW190521 to straddle the lower and upper bounds of the mass gap \citep{Fishbach_2020}. There has also been work by \citet{Nitz_2021} (hereafter \NC) reanalyzing the parameter estimates of GW190521 with the recently released \textsc{IMRPhenomPXHM} waveform \citep{pratten2020lets} which supports mass ratios $q \equiv m_1/m_2 > 4$ that were not considered in the LVC analysis. NC21 found that GW190521 may be an intermediate mass ratio merger, reporting a multimodal posterior with an additional high mass ratio mode not identified in the LVC analysis. The reported source frame component masses for the high mass ratio mode squarely puts each outside of the theorized mass gap with $m_1^{\mathrm{src}} = 166_{-35}^{+16}\msun$ and $m_2^{\mathrm{src}} = 16_{-3}^{+14}\msun$ \citep{Nitz_2021}. Figure \ref{fig:GWTC2_contours} shows the 90\% contours on the posterior samples from events in GWTC-2 with both the LVC GW190521 posterior samples in addition to samples from \NC{} highlighted. This illustrates how differences in the analysis of GW190521 can considerably change the posterior support for its component masses to lie in the theorized PISNe mass gap. If GW190521 does ``straddle" the gap, it would signal the existence of a high-mass population that could inform questions in both astrophysics and cosmology \citep{Ezquiaga:2020tns}.

In this letter we present a simple phenomenological population model parameterizing the PISNe mass gap that enforces a zero rate of BBH mergers within the gap. Our model is a complementary approach to other physically motivated models that describe the impact of PISNe on the mass spectrum \citeg{baxter2021gap}. Using this model we evaluate the evidence for the presence of a mass gap in LIGO/Virgo's second gravitational wave transient catalog, and constrain its properties. We conduct each analysis twice, first using posterior samples for GW190521 released by the LVC, and alternatively using samples produced in \NC. In Section \ref{sec:methods} we introduce our parameterized mass gap model, and the methods used to infer population properties. In Section \ref{sec:results} we present the results of our inference with both sets of posterior samples and two underlying mass distributions. We then discuss our interpretation of the results and astrophysical implications in Section \ref{sec:discussion} and finish with our conclusions on the support for the presence of an upper mass gap in LIGO/Virgo's BH population in Section \ref{sec:conclusion}. 

\section{Methods} \label{sec:methods}

\begin{table*}[t]
\begin{tabular}{|l|l|l|l|}
\hline
\multicolumn{4}{|c|}{\textbf{Primary Mass Model Parameters}} \\ \hline
\textbf{Model} & \textbf{Parameter} & \textbf{Description} & \textbf{Prior} \\ \hline
\multirow{3}{*}{\textsc{Truncated}} & $\alpha$ & slope of the powerlaw & U(-4, 12) \\ \cline{2-4} 
 & $m_\mathrm{min}$ & minimum mass cutoff (\msun) & U(2\msun, 10\msun) \\ \cline{2-4} 
 & $m_\mathrm{max}$ & maximum mass cutoff (\msun) & 200 \msun \\ \hline
\multirow{6}{*}{\textsc{PowerLaw+Peak}} & $\alpha$ & slope of the powerlaw & U(-4, 12) \\ \cline{2-4} 
 & $m_\mathrm{min}$ & minimum mass cutoff (\msun) & U(2\msun, 10\msun) \\ \cline{2-4} 
 & $m_\mathrm{max}$ & maximum mass cutoff (\msun) & 200 \msun  \\ \cline{2-4} 
 & $\mu_p$ & mean of gaussian peak (\msun) & U(20\msun, 70\msun) \\ \cline{2-4} 
 & $\sigma_p$ & width of the gaussian peak (\msun) & U(0.4\msun, 10\msun) \\ \cline{2-4} 
 & $\lambda_p$ & fraction of BBH in the gaussian component & U(0, 1) \\ \hline \hline
\multicolumn{4}{|c|}{\textbf{Mass Ratio Model Parameters}} \\ \hline
\textsc{PowerLaw MassRatio} & $\beta_q$ & slope of the mass ratio powerlaw & U(-4, 12) \\ \hline \hline
\multicolumn{4}{|c|}{\textbf{Redshift Evolution Model Parameters}} \\ \hline 
\textsc{PowerLaw Redshift} & $\gamma$ & slope of redshift evolution powerlaw $(1+z)^\gamma$ & U(-6, 6) \\ \hline \hline
\multicolumn{4}{|c|}{\textbf{Mass Gap Parameters}} \\ \hline 
\multirow{2}{*}{\textsc{Agnostic MassGap}} & $m_g$ & lower bound of PISNe mass gap (\msun) & U(40\msun, 100\msun) \\ \cline{2-4} 
 & $w_g$ & width of the PISNe mass gap (\msun) & U(0\msun, 160\msun) \\ \hline
\multirow{2}{*}{\textsc{Informed MassGap}} & $m_\mathrm{g, min}$ & lower bound of the PISNe mass gap (\msun) & $\mathcal{N}(\mu=\lowerPISN, \sigma=10\msun)$ \\ \cline{2-4} 
 & $m_\mathrm{g, max}$ & upper bound of the PISNe mass gap (\msun) & $\mathcal{N}(\mu=\upperPISN, \sigma=20\msun)$ \\ \hline
\end{tabular}
\caption{Prior choices and description of hyperparameters for used population models.}
\label{table:priortable}
\end{table*}

\subsection{Hierarchical Inference} \label{sec:heirarchical inference}

We use hierarchical Bayesian inference to simultaneously infer hyperparameters of the population distribution of the primary masses ($m_1$), mass ratios ($q$) and the redshifts ($z$) of observed BBHs. We assume the BBH merger rate $d\mathcal{R}$ over a given interval of masses and redshifts can be factored as:

\begin{equation} \label{number_density}
    \frac{d\mathcal{R}(m_1, q, z | \mathcal{R}_0, \Lambda)}{dm_1dq} = \mathcal{R}_0 p(m_1 | \Lambda) p(q | m_1, \Lambda) p(z | \Lambda)
\end{equation}
\noindent
with $\Lambda$ the population hyperparameters and $\mathcal{R}_0$ the local ($z=0$) merger rate. Under the condition that $p(m_1| \Lambda)$ and $p(q |m_1, \Lambda)$ are both normalized, and $p(z)$ chosen such that $p(z=0)=1$, integrating the merger rate density across all primary massses and mass ratios at a given $z$, returns the total BBH merger rate density at that redshift, $\mathcal{R}(z)$. The number density of BBH mergers can be related to the merger rate density by:

\begin{equation}
     \frac{dN(m_1, q, z | \mathcal{R}_0, \Lambda)}{dm_1dqdz} = \frac{dV_c}{dz}\bigg(\frac{T_\mathrm{obs}}{1+z}\bigg) \frac{d\mathcal{R}(m_1, q, z | \mathcal{R}_0, \Lambda)}{dm_1dq}
\end{equation}
\noindent
with $V_c$ the comoving volume element and $T_\mathrm{obs}$ the total observing time with the factor of $1+z$ converting source-frame time to detector-frame. Integrating the above number density across all primary masses, mass ratios and redshifts out to a maximum $z_\mathrm{max}$ returns the expected number of BBH mergers in the universe out to $z_\mathrm{max}$. Given a set of data $\{d_i\}$ from $N_\mathrm{obs}$ gravitational wave events, we can calculate the posterior on $\Lambda$ following e.g.~\citet{Farr_2019} and \citet{Mandel_2019}:

\begin{widetext}
\begin{equation} \label{posterior}
    p\left(\mathcal{R}_0, \Lambda | \{d_i\} \right) \propto p(\Lambda) p(\mathcal{R}_0) e^{-\mathcal{R}_0 \langle VT \rangle_{\Lambda}} \prod_{i=1}^{N_\mathrm{obs}} \Bigg[ \int \mathcal{L}\left(d_i | m_1^i, q^i, z^i \right) \frac{d\mathcal{R}}{dm_1 dq dz}(\Lambda) dm_1 dq dz \Bigg],
\end{equation}
\noindent
where $\mathcal{L}(d_i|m_1, q, z)$ is the single-event likelihood function used to infer each event's parameters, and $\langle VT \rangle_{\Lambda}$ is the average sensitive time-volume when assuming a population corresponding to hyper-parameters $\Lambda$. To estimate the $\langle VT \rangle_{\Lambda}$, we use the results of the LVC's injection campaign where the GWs from a fixed, broad population of sources were simulated, injected into real detector data, and searched for using the same analyses that produced GWTC-2 \footnote{For O3a we used the injection sets used by \citet{o3a_pop}, which can be found at https://dcc.ligo.org/LIGO-P2000217/public. For O1/O2 we used the mock injection sets used by \citet{o1o2_pop} which can be found at https://dcc.ligo.org/LIGO-P2000434/public}.  We use importance sampling over the detected simulated events to estimate $\langle VT \rangle_{\Lambda}$, marginalizing over the uncertainty in our estimate due to a finite number of simulated events, following \cite{Farr_2019}. We assume that repeated sampling of $\langle VT \rangle_{\Lambda}$ will follow a normal distribution (i.e. $\langle VT \rangle_{\Lambda} \sim \mathcal{N}(\mu(\Lambda), \sigma(\Lambda))$), with $\mu$ the raw importance sampled estimate and $\sigma$ standard error. Now we define $N_\mathrm{eff}$, the effective number of independent draws contributing to the importance sampled estimate, as $N_\mathrm{eff} \equiv \frac{\mu^2}{\sigma^2}$, which we verify to be sufficiently high after re-weighting to a population (i.e. $N_\mathrm{eff} > 4N_\mathrm{det}$). After marginalizing over the uncertainty estimating the sensitive time-volume, we write the marginalized posterior as:

\begin{equation} \label{xi-marged-posterior}
    p\left(\mathcal{R}_0, \Lambda | \{d_i\} \right) \propto  \\
    p(\Lambda) p(\mathcal{R}_0) \prod_{i=1}^{N_\mathrm{det}} \Bigg[ \int \mathcal{L}\left(d_i | m_1^i, q^i, z^i \right) p(m_1^i, q^i, z^i | \Lambda) dm_1 dq dz \Bigg] \mathcal{R}_0^{N_\mathrm{obs}} \exp{\frac{\mathcal{R}_0\mu(\mathcal{R}_0\mu - 2N_\mathrm{eff})}{2N_\mathrm{eff}}}
\end{equation}
\noindent
Finally, when using the commonly chosen log-Uniform prior on $\mathcal{R}_0$ \citep{o3a_pop}, we can marginalize over the local merger rate, neglecting terms of $\mathcal{O}(N_\mathrm{eff}^{-2})$ or greater: \citep{Farr_2019}

\begin{equation}\label{importance-posterior}
    \log p\left(\Lambda | \{d_i\}\right) \propto \sum_{i=1}^{N_\mathrm{obs}} \log \bigg[ \frac{1}{K_i} \sum_{j=1}^{K_i} \frac{p(m_1^{i,j}, q^{i,j}, z^{i,j} | \Lambda)}{\pi(m_1^{i,j}, q^{i,j}, z^{i,j})} \bigg] -  \\
    N_\mathrm{obs} \log \mu + \frac{3N_\mathrm{obs} + N_\mathrm{obs}^2}{2N_\mathrm{eff}} + \mathcal{O}(N_\mathrm{eff}^{-2})
\end{equation}

%\begin{equation} \label{R-marged-posterior}
%    \log p\left(\Lambda | {d_i}\right) \propto \sum_{i=1}^{N_\mathrm{obs}} \int \log \mathcal{L}\left(d_i %| m_1^i, q^i, z^i \right) p(m_1^i, q^i, z^i | \Lambda) dm_1 dq dz -  \\
%    N_\mathrm{obs} \log \mu + \frac{3N_\mathrm{obs} + N_\mathrm{obs}^2}{2N_\mathrm{eff}} 
%\end{equation}

\end{widetext}

\noindent
In the last expression we further approximated the inner integral over the individual event parameters $m_1^i, q^i, z^i$ with importance sampling over $K_i$ single-event posterior samples generated from inference with prior $\pi(m_1^{i,j}, q^{i,j}, z^{i,j})$. To calculate marginal likelihoods and draw samples of the hyper parameters from the hierarchical posterior distribution shown in equation \ref{importance-posterior}, we use the \textsc{Bilby} \citep{Ashton_2019, bilby_gwtc1} and \textsc{GWPopulation} \citep{Talbot_2019} Bayesian inference software libraries with the \textsc{Dynesty} dynamic nested sampling algorithm \citep{Speagle_2020}.

% \subsection{Model Comparison}\label{model-comparison}

% To compare competing models in the aforementioned Bayesian framework we calculate Bayes factors \footnote{The true ``Bayesian" way to compare models is using odds ratios, which are Bayes factors multiplied by the ratio of prior odds of each model. Because we don't a priori have expectations of which population model would be more likely, Bayes factors are odds ratios with equal prior odds for each model}. 

% \begin{equation} \label{BF}
%  BF_{B}^{A} \equiv \frac{\mathcal{Z}_{A}(\{d_i\})}{\mathcal{Z}_{B}(\{d_i\})}
% \end{equation}
% \begin{equation}\label{Z}
%  \mathcal{Z}(\{d_i\}) = \int d\Lambda p(\Lambda | \{d_i\}) 
% \end{equation}

% The Bayes factor in Eq.\ref{tab:BF} is the ratio of marginal likelihoods of two competing models, A and B, with the marginal likelihood defined in Eq.\ref{Z}. $BF_{B}^{A} > 1$ indicates that model A is preferred. To determine the strength of model comparisons, $0 < \ln BF < 2$ indicates inconclusive preference, $2 < \ln BF < 6$ indicates slight preference, and $\ln BF > 6$ indicates strong preference.

\begin{figure*}
    \centering
    \includegraphics[width=0.9\textwidth]{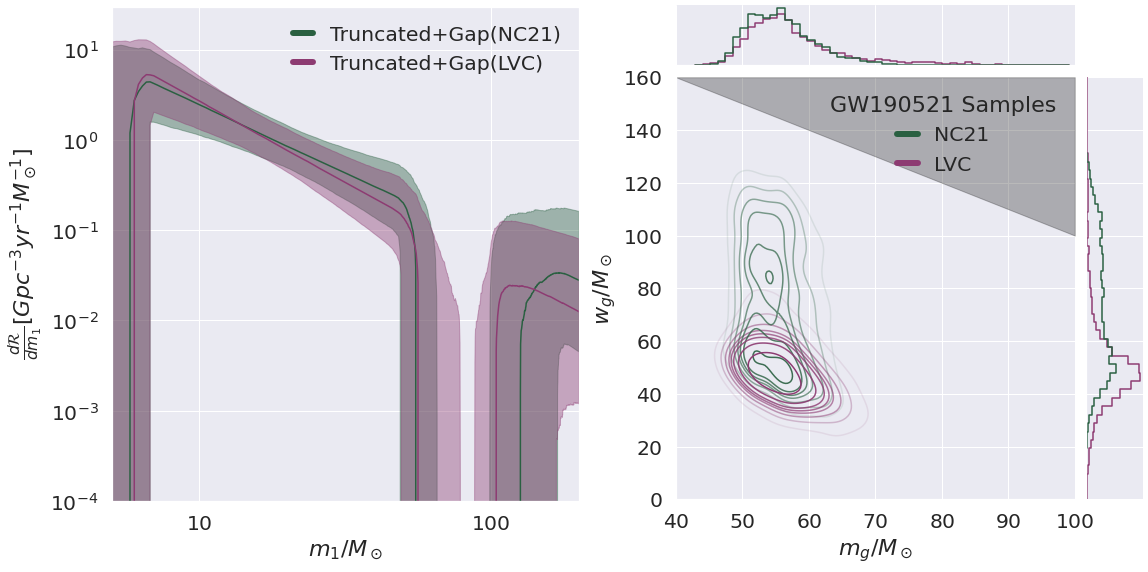}
    \caption{Posterior merger rate density (left) as function of primary mass inferred with the mass gap imposed on top of the \textsc{Truncated} model from \citet{o3a_pop}. Solid curves shown the median posterior sample, while the shaded regions show the 90\% credible level. 1-d and 2-d marginal posterior samples (right) of the two mass gap parameters, the lower edge, $m_g$, and the width, $w_g$, both with uniform agnostic priors over the range shown. The contour lines enclose 10-80\% of the posterior. The grey region shows where our model reduces to the \textsc{Truncated} model with maximum mass at $m_g$. Results are shown using both the GW190521 posterior samples reported by the LVC (purple) and \citet{Nitz_2021} (green).}
    \label{fig:dRs_truncated}
\end{figure*}

\subsection{Parameterized Mass Gap Model} \label{sec:massgap model}

We build on models used in \citet{o3a_pop}, \citet{o1o2_pop}, and \citet{Fishbach_2018redshift} for the mass ratio and redshift of our population, specifically $p(z | \gamma) \propto (1+z)^\gamma$ and $p(q | m_1, m_\mathrm{min}, \beta_q) \propto q^{\beta_q} \Theta(qm_1 - m_\mathrm{min}) \Theta(m_1 - qm_1)$ with $\Theta$ denoting the Heaviside step function ensuring $m_2$ is within the range [$m_\mathrm{min}$, $m_1$]. We choose to neglect a population model for the spins, assuming that their population follows the uniform and isotropic prior used in each event's initial analysis. For the primary mass distribution we use two different models presented in \citet{o3a_pop}, the \textsc{Truncated} and \textsc{PowerLaw+Peak} models. We choose to build upon the \textsc{Truncated} model for its simplicity and the \textsc{Powerlaw+Peak} model since \citet{o3a_pop} found it to have the highest marginal likelihood of the models used. Additionally, it is important to include the peak (\textsc{Powerlaw+Peak}), as it was motivated to model the pileup of events due to PPISN mass loss which is expected from the same processes predicting the upper mass gap \citep{Talbot_2018}. We then introduce a mass gap into both the primary and mass ratio distributions by enforcing that neither component mass can lie within the gap, which we parameterize with the location of the lower edge $m_g$ and the width of the gap $w_g$:

\begin{equation*}
    \label{eq:m1dist}
    p(m_1 | m_g, w_g, \Lambda) \propto \begin{cases}
    0 & m_g \leq m_1 \leq m_g + w_g \\
    p(m_1 | \Lambda) & \text{otherwise}

    \end{cases}
\end{equation*}

\begin{equation*}
    \label{eq:qdist}
    p(q | m_g, w_g, m_1, \Lambda) \propto \begin{cases}
    0 & m_g \leq m_1q \leq m_g + w_g \\
    p(q | m_1, \Lambda) & \text{otherwise}
    \end{cases}
\end{equation*}

Our model prescribes a zero-rate within the mass gap, which might be expected if the entire population of sources is formed through stellar collapse. We enforce an overall maximum BH mass of 200\msun~so that if the upper edge of the gap is not constrained (i.e. $m_g+w_g \geq 200\msun$), it is equivalent to the underlying primary mass model with a maximum mass at $m_g$. The hyperparameters' descriptions according to each of the models used along with their corresponding priors can be found in Table \ref{table:priortable}. 

\subsection{Injections and Sensitivity Estimates} \label{sec:sensitive_inj}

The injection sets reported and used by the LVC in \citet{o3a_pop} only include simulated signals with source frame masses up to 100\msun. However, since \NC{} found GW190521 to have posterior support for its primary source frame mass to be up to 180\msun, we want to probe this region of parameter space. To prevent inferring an artificially high merger rate above the gap, our mass gap models are chosen to enforce the same powerlaw index in the region above the gap as below. This fixes the normalization above the gap based on the powerlaw fit below the gap that is influenced by the events which have $m_1 < 60\msun$, which is the majority. This is also in-line with the expectation that very massive BHs can also be produced through stellar evolution, and thus come from the same stellar population as below the gap \citep{Renzo_2020}. We additionally fit our population models with the same injection set truncated to only include injections with masses up to 80\msun, and found that it did not bias our results.

\section{Results} \label{sec:results}

We fit our population models to the 46 definitive BBH mergers in GWTC-2 (i.e., excluding GW170817, GW190425, GW190814, and GW190426\_15215 \citep{170817disc, 190425disc, 190814disc}), and since we are focused on the details of the high-mass population we neglect the low-mass smoothing feature of the models used in \citet{o3a_pop} for simplicity. Fig.~\ref{fig:dRs_truncated} (left) shows the inferred merger rate density as a function of primary mass, when using the gap model on top of the underlying \textsc{Truncated} model, in which there is clear inference of a mass gap. When using the \textsc{Truncated} model, the choice of GW190521 posterior samples (either from the LVC or \NC) does not significantly affect the outcome, but the gap is inferred to be narrower and with an upper edge at lower masses when using the LVC samples. Fig \ref{fig:dRs_truncated} (right) shows the 1-d and 2-d marginal posterior distributions for the two gap parameters. Here we can see that the posterior distribution for the width of the gap is less constrained using \NC{} samples, but both cases show little support for a zero-width gap. Using the LVC GW190521 samples we find support for the \textsc{Truncated+Gap} over \textsc{Truncated} model with Bayes factor $\ln \mathcal{B} = 2.79$, with lower and upper bounds at $55.12_{-4.38}^{+7.54} \msun$, and $103.74_{-6.32}^{+17.01}\msun$ respectively. While the gap model is clearly favored in that comparison, when using samples from \NC's for GW190521 the gap is more clearly favored with $\ln \mathcal{B} = 6.5$, with lower and upper bounds at $55.33_{-4.21}^{+5.21} \msun$, and $126.03_{-22.65}^{+30.25}\msun$.

\begin{figure*}
    \centering
    \includegraphics[width=0.9\textwidth]{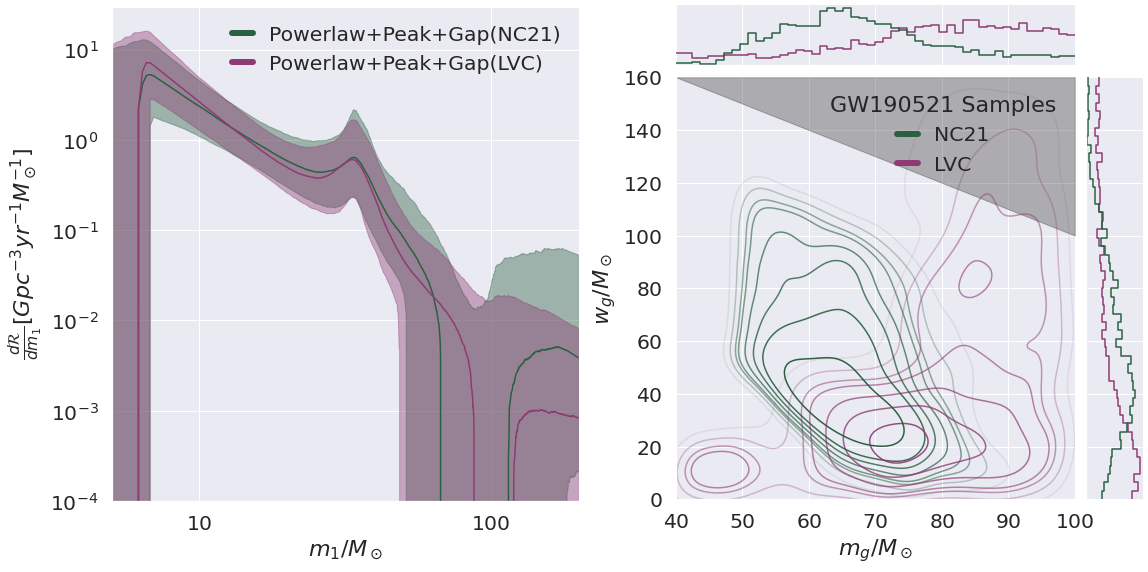}
    \caption{Posterior merger rate density (left) as function of primary mass inferred with the mass gap imposed on top of the \textsc{Powerlaw+Peak} model from \citet{o3a_pop}. Solid curves shown the median posterior sample, while the shaded regions show the 90\% credible level. 1-d and 2-d marginal posterior samples (right) of the two mass gap parameters, the lower edge, $m_g$, and the width, $w_g$, both with uniform agnostic priors over the range shown. The contour lines enclose 10-80\% of the posterior. The grey region shows where our model reduces to the \textsc{Powerlaw+Peak} model with maximum mass at $m_g$. Results are shown using both the GW190521 posterior samples reported by the LVC (purple) and \citet{Nitz_2021} (green).}
    \label{fig:dRs_plpeak}
\end{figure*}

Fig \ref{fig:dRs_plpeak} (left) shows the inferred merger rate density as a function of primary mass when imposing a gap on the most favored mass model in \citet{o3a_pop}, \textsc{Powerlaw+Peak}. When including the Gaussian peak in our primary mass distribution, the support for the upper mass gap significantly reduces, regardless of which GW190521 samples are used. We find log-Bayes factors for inclusion of the gap to be $\ln \mathcal{B} = -0.5$ and $\ln \mathcal{B} = 0.5$ when using the LVC, and \citet{Nitz_2021} GW190521 posterior samples respectively.  Fig.~\ref{fig:dRs_plpeak} (right) shows the 1-d and 2-d marginal posterior distributions for the gap parameters, which show poorer constraints on the gap in the \textsc{Powerlaw+Peak+Gap} model relative to \textsc{Powerlaw+Gap}. In this case, both choices of posterior samples show support for a zero-width gap, as reflected in the Bayes factors. 

\section{Discussion} \label{sec:discussion}

\begin{table}[t]
\begin{tabular}{l|r|r}
Model & LVC & \NC{} \\ \hline
\textsc{Truncated} & -4.98  & -7.99 \\
\textsc{Truncated+Gap} & -2.20 & -1.51 \\
\textsc{Truncated+Gap} (informed) & -0.87 & 0.0 \\
\textsc{Powerlaw+Peak} & 0.0 & -1.93 \\
\textsc{Powerlaw+Peak+Gap} & -0.57 & -1.35 \\
\textsc{Powerlaw+Peak+Gap} (informed) & -1.05 & -0.95 \\
\end{tabular}
\caption{Log Bayes factors for the models analyzed in this work, shown relative to the most favored model in each column. The two columns show results with the LVC reported GW190521 parameter estimation samples vs. those reported by \NC{}.}
\label{tab:BF}
\end{table}

Our results are inconclusive about the existence of a high-mass mass gap.  While a gap is clearly inferred when using a pure power-law model of the population, adding a Gaussian peak to the mass distribution washes away the need for the gap. Furthermore, differences in parameter estimates with different priors and waveforms gives rise to different inferences on the gap parameters if a gap indeed exists. These results are summarized in Table \ref{tab:BF} through Bayes factors comparing the marginal likelihood of each model to that of the model with highest marginal likelihood (which therefore has $\ln \mathcal{B} = 0$). The Bayes factors for LVC and \NC{} parameter estimates are treated separately in the table. We also include Bayes factors for analyses with ``informed" priors on the gap boundary, where we place Gaussian priors for $m_\mathrm{g, min}$ and $m_\mathrm{g, max}$ around centered on the approximate expected gap bounds (i.e. $p(m_\mathrm{g, min}) \sim \mathcal{N}(\mu=\lowerPISN, \sigma=10\msun)$ and $p(m_\mathrm{g, max}) \sim \mathcal{N}(\mu=\upperPISN, \sigma=20\msun)$). With the smaller prior volume in these runs, the Bayes factors are higher than with the uninformed gap priors. Nevertheless, these Bayes factors do not increase enough to change the general finding of this work that the gap is favored with the \textsc{Truncated} model, but its existence is unclear when considering the \textsc{Powerlaw+Peak} model.

While the gap (if it exists) is difficult to resolve at present due to low number statistics, future detections will enable a finer look at the gap. Three extensions should be made when more high-mass detections are available, which we have eschewed for now due to the single event GW190521 driving the inference:
\begin{enumerate}
    \item Allow a different mass ratio distribution for high-total-mass events than for low-total-mass.
    \item Allow a non-zero rate in the gap, possibly with spins enforced to be near $\chi\sim0.7$ to account for hierarchical mergers \citep{Gerosa:2017kvu,Fishbach:2017dwv,Kimball_genealogy,doctor2021black}.
    \item Allow the merger rate normalization above the gap to be a free parameter.\footnote{This was not possible in this work due to LVC injections only reaching source frame component masses of $100\,M_\odot$, making the overall rate above that threshold unconstrained, as was discuessed in Section \ref{sec:sensitive_inj}}% Hence we assumed the normalization above the gap follows that below the gap.}
\end{enumerate}

The location of the lower edge of the PISNe mass gap has been found to be insensitive to many variations in stellar physics, especially metallicities \citep{Farmer_2019}. A metallicity independent feature in the BH mass spectrum can provide a ``standard siren" that allows for independent measurements of redshift and luminosity distances to GW sources to directly measure the Hubble constant \citep{Farr_2019HUB}. The lower edge of the PISNe mass gap has been found to be very sensitive to variations in the ${}^{12}\mathrm{C}(\alpha, \gamma)^{16}\mathrm{O}$ reaction rate, with some choices of rate pushing the lower bound up to $\sim 90\msun$, illustrating that constraints on the lower bound can also be used to put constraints on nuclear physics going on inside stars' cores. \citep{Farmer_2019, Farmer_2020}. These astrophysical implications rely on the CO-BH mass relation from \citet{Farmer_2019} that predicts a pileup of BHs below the onset of the PISNe mass gap, implying constraints on the gap lower bound that neglect a pileup could not be reliably used to constrain nuclear physics. The upper edge of the mass gap, is currently not well constrained, but \citet{Ezquiaga:2020tns} argues LIGO/Virgo (at A+ sensitivity) will be sensitive to BBH's with component masses that could lie above the PISNe gap. Future constraints on the upper edge may also provide a novel probe of physics beyond the standard model \citep{Croon_newphysics}.

\section{Conclusions}\label{sec:conclusion}
Black holes formed through stellar collapse are expected to have a gap in their mass spectrum from $\sim\lowerPISN-\upperPISN$. We assess the support, or lack thereof, for the existence of such a gap in the GWTC-2 catalog, using two parameterized black hole binary merger population models.  Our population models build on the \textsc{Truncated} and \textsc{Powerlaw+Peak} models previously fit to these catalogs, and explicitly allow for a zero-rate mass gap with a population of black holes above the gap. Our analyses also consider two separate inferences of GW190521 parameters, one from the LVC and the other from \citet{Nitz_2021}. We find that the results of our inference regarding the existence of a gap are contingent in part on the choice of population model and GW190521 parameter estimation results.  

If a pure power law is used to describe the distribution of primary masses, we infer a mass gap from $56.12_{-4.38}^{+7.54} \msun$ to $103.74_{-6.32}^{+17.01}\msun$, however if the data support more unequal masses for GW190521 as suggested in \citet{Nitz_2021}, we infer a mass gap from $55.33_{-4.21}^{+5.21} \msun$ to $126.03_{-22.65}^{+30.25}\msun$. When using a power law with an additional Gaussian component, we no longer find significant support for a zero-rate mass gap. This does not, however, imply the nonexistence of a mass gap due to PISNe but points towards there being a secondary population of BHs that LIGO/Virgo is observing not formed through isolated stellar evolution. Future studies may be able to distinguish between these multiple formation channels in part by looking for a zero-rate gap in BH sub populations while additionally using informed constraints on expected properties that a hierarchically formed population of BHs would have.

\section{Acknowledgements}\label{sec:acknowledments}

This research has made use of data, software and/or web tools obtained from the Gravitational Wave Open Science Center (\url{https://www.gw-openscience.org/}), a service of LIGO Laboratory, the LIGO Scientific Collaboration and the Virgo Collaboration. This work benefited from access to the University of Oregon high performance computer, Talapas.  This material is based upon work supported in part by the National Science Foundation under Grant PHY-1807046 and work supported by NSF’s LIGO Laboratory which is a major facility fully funded by the National Science Foundation.

\software{
\textsc{Astropy}~\citep{2018AJ....156..123A},
\textsc{NumPy}~\citep{harris2020array},
\textsc{SciPy}~\citep{2020SciPy-NMeth},
\textsc{Matplotlib}~\citep{Hunter:2007},
\textsc{seaborn}~\citep{waskom2020seaborn},
\textsc{bilby}~\citep{Ashton_2019},
\textsc{GWPopulation}~\citep{Talbot_2019}
}

\bibliography{references}{}
\bibliographystyle{aasjournal}

\end{document}